\documentclass[conference]{IEEEtran}
\IEEEoverridecommandlockouts

\usepackage{cite}
\usepackage{graphicx}
\usepackage{grffile}
\usepackage{amsmath,amssymb}
\usepackage{booktabs}
\usepackage{multirow}
\usepackage{array}
\usepackage{tabularx}
\usepackage{adjustbox}
\usepackage{algorithm}
\usepackage[noend]{algpseudocode}
\usepackage{subcaption}
\usepackage{xcolor}
\usepackage{xurl}
\usepackage{enumitem}
\usepackage{url}

\newcommand{\pkO}{\mathsf{pk}_{\mathcal O}}
\newcommand{\skO}{\mathsf{sk}_{\mathcal O}}
\newcommand{\Low}{\textsf{Low}}
\newcommand{\Mod}{\textsf{Moderate}}
\newcommand{\High}{\textsf{High}}

\title{Smart Blockchain-Based Access Control for the Internet of Things}
\author{
  \IEEEauthorblockN{1\textsuperscript{st} Mahdi Manavi}
  \IEEEauthorblockA{
    \textit{University of Houston}\\
    Houston, USA \\
    Mmanavi@central.uh.edu}
  \and
  \IEEEauthorblockN{2\textsuperscript{nd} Yunpeng Zhang}
  \IEEEauthorblockA{
    \textit{University of Houston}\\
    Houston, USA\\
    yzhan226@central.uh.edu}
  \and
  \IEEEauthorblockN{3\textsuperscript{rd} Guoning Chen}
  \IEEEauthorblockA{
    \textit{University of Houston}\\
    Houston, USA \\
    gchen22@central.uh.edu}
}

\newcommand{\incfig}[2]{%
  \begingroup\edef\__figfile{\detokenize{#1}}%
  \IfFileExists{\__figfile}{%
    \includegraphics[width=#2]{\__figfile}%
  }{%
    \fbox{\parbox[c][0.18\textheight][c]{#2}{\centering Missing figure: \texttt{\__figfile}}}%
  }%
  \endgroup%
}

\begin{document}
\maketitle

\begin{abstract}
Securing access control in large-scale Internet of Things (IoT) deployments requires mechanisms that adapt to risk while preserving low latency for benign traffic. Permissioned blockchains such as Hyperledger Fabric offer auditability through smart contracts, but static endorsement policies impose the same validation depth on all requests, regardless of security posture. We propose a risk-adaptive enforcement layer for Hyperledger Fabric that couples an off-chain LSTM-based risk oracle with deterministic on-chain checks. The oracle assigns each request to a tier (\Low, \Mod, \High) and issues a signed attestation bound to the client identity and target key/version. Endorsing peers verify the attestation in chaincode and enforce tier-conditioned SBE policies without modifying the ordering service or consensus. Experiments on a Fabric testbed show that tier-conditioned endorsement strengthens validation for higher-risk requests while retaining low confirmation latency for benign workloads.
\end{abstract}


\section{Introduction}
\IEEEPARstart{I}{oT} deployments combine massive device heterogeneity with stringent latency targets, making access control essential yet hard to scale in practice~\cite{IEEEhowto:jadid11}. The growing reliance on cloud-hosted control planes expands the attack surface, while resource constraints at the edge amplify the impact of abuse and disruption~\cite{Manavi2023SmartCloud}. Anomaly-based detection is particularly useful for previously unseen (e.g., zero-day) behaviors~\cite{Manavi2019GRUGA}, but even accurate detection must be coupled with enforceable, low-latency access-control decisions. Classical access-control models (role-, credential-, and trust-based) are widely used~\cite{IEEEhowto:RXu,IEEEhowto:Duk}, yet centralized decision points introduce single points of failure and limit scalability under churn and adversarial pressure~\cite{IEEEhowto:Raibi}.

A key challenge is that request \emph{risk} varies substantially across devices and over time, yet most systems apply a fixed validation depth to all requests—penalizing benign traffic while under-scrutinizing suspicious bursts. This motivates risk-adaptive access control that selectively adjusts enforcement effort per request while remaining auditable and operationally simple~\cite{NIST-RAdAC,NIST-Glossary-RAdAC}. IoT telemetry also introduces practical constraints (noisy features, tight QoS), encouraging designs that keep enforcement lightweight and deterministic while pushing heavier analytics off-chain~\cite{Manavi2024QoS,Manavi2024MOGAMV}.

Blockchain has been explored as a decentralized, tamper-evident substrate for IoT access control~\cite{IEEEhowto:shak,IEEEhowto:hao}. While smart contracts improve auditability~\cite{IEEEhowto:mohanta}, conventional designs often impose latency that conflicts with real-time IoT requirements~\cite{IEEEhowto:rustemi}, and many prior ``adaptive'' strategies tune parameters to network load rather than per-request security posture.

We address these gaps with a \emph{risk-adaptive blockchain framework} integrating machine learning with a permissioned ledger. An off-chain LSTM-based oracle maps each request to a three-tier label (\Low/\Mod/\High) returned as a signed attestation. On-chain, endorsing peers deterministically verify the attestation and enforce a tier-conditioned \emph{State-Based Endorsement (SBE)} profile, preserving Fabric's execute--order--validate pipeline with no modification to the ordering service~\cite{Androulaki2018,SBE-Sample}. The main contributions are:
\begin{itemize}
  \item A Fabric-compatible, risk-adaptive enforcement workflow separating off-chain learning from on-chain deterministic policy checks.
  \item A tier-conditioned mapping from oracle-attested risk levels to per-key SBE profiles, enabling fine-grained endorsement hardening without changing consensus.
  \item A reproducible experimental study comparing tier-conditioned SBE against static endorsement baselines, quantifying latency/throughput trade-offs and operational considerations.
\end{itemize}

\section{Related Work}
\label{sec:relatedwork}

Blockchain-backed access control for IoT has been widely studied to support decentralized trust, auditability, and cross-organization policy enforcement~\cite{bobde,IEEEhowto:Novo,IEEEhowto:Pinno,IEEEhowto:Ding1}. Many proposals encode authorization logic in smart contracts, leveraging the ledger for immutable logging and dispute resolution~\cite{IEEEhowto:Novo,IEEEhowto:Pinno,IEEEhowto:Ding1}, but typically apply \emph{static} validation effort once deployed. Risk-adaptive access control (RAdAC) varies enforcement based on runtime context and estimated threat~\cite{NIST-RAdAC,NIST-Glossary-RAdAC}, yet remains underexplored in practical permissioned-blockchain pipelines. In Hyperledger Fabric, endorsement policies and SBE can require additional co-signers for sensitive updates but do not by themselves determine \emph{when} stricter endorsement should apply~\cite{Androulaki2018,SBE-Sample}.

Prior work has explored static policy hardening, attribute-based access control (ABAC), and external access gateways for Fabric deployments; however, integrating dynamic risk signals directly into the endorsement layer remains less explored. A practical challenge is determinism: risk estimation uses probabilistic models and evolving histories, while Fabric endorsers must execute chaincode deterministically so that identical proposals yield identical read--write sets. Our design addresses this constraint by keeping risk scoring off-chain and using only compact, signed attestations on-chain, aligning with the principle that complex analytics run off-chain while on-chain logic enforces simple, auditable checks. Another concern is policy churn: naively updating per-key SBE on every request may create additional metadata writes and increase MVCC conflicts; our two-phase rotation (Sec.~\ref{subsec:sbe-path}) separates policy updates from functional writes to avoid this. The proposed approach is also complementary to network-level defenses (rate limiting, mTLS): those reduce volumetric load, while tier-conditioned SBE improves \emph{integrity} by requiring stronger cross-organization validation for higher-risk operations.

\section{Proposed Approach}
\label{sec:proposed}

\subsection{System Overview}

\begin{figure*}[h]
\centering
\incfig{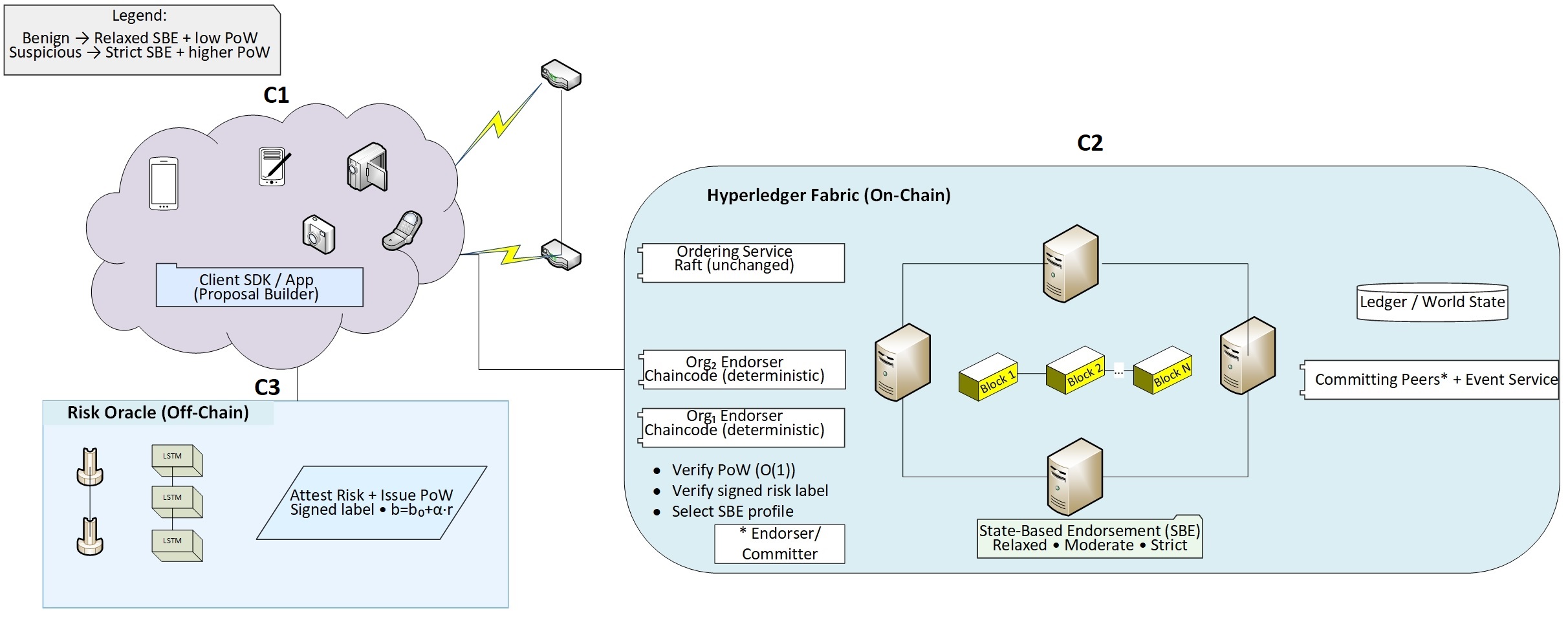}{7.3in}
\caption{System overview. The risk oracle runs off-chain and returns a \emph{signed} risk label to the client. The client attaches the signed label to the Fabric proposal. Endorsing peers verify it deterministically in chaincode and select an SBE profile; ordering (Raft) remains unchanged.}
\label{fig_sim}
\end{figure*}

Figure~\ref{fig_sim} summarizes the three-layer architecture: (1) IoT clients, (2) an off-chain risk oracle, and (3) on-chain execution on Hyperledger Fabric. Each client contacts the oracle before submitting a Fabric proposal. The oracle returns a \emph{signed} risk label bound to the client identity and target key/version. The client attaches the attestation to its Fabric proposal. Endorsing peers verify the attestation deterministically in chaincode and select an SBE profile whose endorsement depth depends on the attested tier. The ordering service (Raft) and its batching policy remain unchanged; all adaptivity is realized through per-key SBE. Benign requests traverse a lightweight path (fewer endorsers), whereas suspicious ones incur stricter multi-endorser checks, preserving blockchain determinism and auditability.

\subsection{Client and Gateway Submission Layer}
\label{subsec:c1-client}

The client layer prepares each access request for Fabric while keeping the blockchain side deterministic. For every operation, the client (or a trusted gateway) collects request metadata---client identity, operation type, target key/namespace, and timestamps---and queries the off-chain oracle to obtain a signed tier attestation. The attestation binds the request to a unique identifier (\texttt{att\_id}) and freshness parameters (expiry/epoch), enabling endorsers to deterministically reject stale or replayed proofs. The client then attaches the oracle-signed attestation to the Fabric proposal and collects endorsements according to the tier-conditioned SBE profile for the target key/namespace.

Resource-constrained devices may delegate proposal construction to a trusted gateway while the gateway preserves the original device identity in request metadata. In this deployment mode, the gateway handles request formatting and oracle communication but cannot alter the tier without invalidating the oracle signature, preserving auditability while reducing direct blockchain overhead on constrained devices.

\subsection{Applying SBE Without Ambiguity}
\label{subsec:sbe-path}

To ensure tier-dependent SBE never introduces race conditions, endorsing peers follow a fixed verification path. Given a proposal $(\mathit{req},\, \ell,\, \sigma)$, the chaincode: (i) verifies the oracle signature, (ii) checks binding to the submitting identity and target key/version, (iii) checks freshness (epoch/expiry) and replay status, and (iv) enforces the SBE policy bound to the attested tier $T$. Algorithm~\ref{alg:endorser} summarizes this path.

\begin{algorithm}
\caption{Chaincode verification (endorsers)}
\label{alg:endorser}
\begin{algorithmic}[1]
\Require proposal $(\mathit{req},\ell,\sigma)$
\State \textbf{VerifySig}$(pk_{\mathrm{orc}},\sigma,\ell)$
\State \textbf{CheckBind}$(\ell, \textit{clientID}, \textit{key/version})$
\State \textbf{EpochFresh}$(\ell)$
\State \textbf{NotSpent}$(\ell.\texttt{att\_id})$
\State \textbf{enforce} SBE policy bound to $\ell.T$
\State \textbf{execute} state mutation under the active policy
\end{algorithmic}
\end{algorithm}

\noindent\textbf{Invariant A (no same-transaction policy change-and-use).}
A transaction must not \emph{both} modify the per-key SBE policy and rely on that modified policy in the same proposal. We enforce this via \emph{two-phase rotation}: (i) a metadata-only \texttt{UpdateSBE} for key $K$ installs the profile selected by oracle label $\ell$, and (ii) functional writes to $K$ occur only \emph{after} receiving the commit event of (i). The write path in Algorithm~\ref{alg:endorser} therefore always executes under an SBE configuration that is already committed on-chain, avoiding TOCTOU inconsistencies and policy-downgrade races.

\noindent\textbf{Anti-replay.}
Each attestation carries a context-bound identifier $\ell.\texttt{att\_id}$ computed as $H(\texttt{epoch} \parallel \texttt{clientID} \parallel \texttt{key/version} \parallel \texttt{nonce})$, binding the attestation to a specific requester and target. The chaincode maintains a spent registry keyed by $\ell.\texttt{att\_id}$; upon a successful commit the identifier is recorded and subsequent proposals carrying the same identifier are deterministically rejected.

\subsection{Off-chain Risk Oracle and Attestation}
\label{subsec:oracle}

The oracle keeps all non-deterministic learning and history aggregation off the blockchain. It ingests request metadata (operation type, key/namespace, client identity, temporal/behavioral features), combines it with a compact historical trust state, and evaluates risk using an LSTM model. The blockchain never invokes the oracle at runtime; it only verifies the oracle's attestation carried in proposals.

\subsubsection{Risk Scoring and Tiering}
\label{subsec:oracle-risk}

Let $y \in \{0,1\}$ denote the LSTM's binary output ($y{=}1$ indicates suspicious), and let $h \in \{\textsf{benign}, \textsf{suspicious}\}$ be the node's historical trust state computed off-chain from rolling acceptance/rejection patterns. The tier is:
\[
T(y,h)=
\begin{cases}
\Low      & \text{if } y{=}0 \land h{=}\textsf{benign},\\
\Mod      & \text{if } (y{=}1 \land h{=}\textsf{benign}) \lor (y{=}0 \land h{=}\textsf{suspicious}),\\
\High     & \text{if } y{=}1 \land h{=}\textsf{suspicious}.
\end{cases}
\]
Consistently benign nodes stay on \Low; requests that are both currently suspicious \emph{and} historically problematic receive \High.

\subsubsection{Oracle Attestation Format}
\label{sec:attestation-format}

The oracle returns a signed label $(\ell,\sigma)$ verifiable deterministically by endorsers:
\begin{equation}
\texttt{att\_id} = H(\texttt{epoch} \parallel \texttt{clientID} \parallel \texttt{key/version} \parallel \texttt{nonce})
\label{eq:att-id}
\end{equation}
\begin{equation}
\ell = \big\langle \mathtt{clientID},\ \mathtt{key/version},\ T,\ \mathtt{epoch},\ \mathtt{att\_id},\ \mathtt{expiry} \big\rangle
\label{eq:ell}
\end{equation}
with $\sigma=\mathrm{Sign}(\skO, \ell)$, where $\pkO$ is distributed via channel configuration. Endorsers verify: (a) $\mathrm{VerifySig}(\pkO,\ell,\sigma)$; (b) binding of $\ell.\texttt{clientID}$ and $\ell.\texttt{key/version}$ to the proposal; (c) freshness ($\text{now} \le \ell.\texttt{expiry}$); (d) $\texttt{att\_id}$ is unspent.

\subsection{On-chain Enforcement and Risk-aware SBE}

Endorsing peers receive Fabric proposals bundling the request and a signed risk label $(\ell,\sigma)$. Each endorser deterministically verifies the attestation and enforces the SBE profile bound to the attested tier:
\begin{equation}
\textsf{Low} \mapsto \textsf{Relaxed},\quad
\textsf{Moderate} \mapsto \textsf{Moderate},\quad
\textsf{High} \mapsto \textsf{Strict}.
\end{equation}
The SBE policy update is always committed in a separate transaction before any state mutation that depends on it. The ordering service (Raft) and consensus semantics remain unchanged.

\subsection{Operational Example}
\label{subsec:example}

Consider a sensor that normally submits periodic status updates to a shared key class. If the LSTM output is benign and the device history is also benign, the oracle signs a \Low-tier attestation and the proposal follows the \emph{Relaxed} SBE path, minimizing latency for routine traffic. If the same device suddenly issues an unusual operation or deviates from its historical pattern, the oracle assigns \Mod. The request is still allowed to proceed, but it must satisfy a stronger endorsement profile requiring additional peer co-signatures. Finally, if both the current request and the historical trust state are suspicious, the oracle assigns \High, forcing the \emph{Strict} cross-organization profile before the state update can commit.

This example illustrates that the system does not simply block anomalous requests at the gateway. Instead, it uses risk to adjust the amount of independent validation required by the ledger. This is particularly useful in consortium IoT settings where a request may be suspicious but still operationally necessary---such as emergency actuation, recovery actions, or rare maintenance operations. The stricter endorsement profile provides accountability and cross-organization agreement without requiring the orderer or consensus layer to distinguish risk tiers directly.

\subsection{Design Rationale}
\label{subsec:rationale}

The design deliberately places the oracle outside the blockchain boundary for three reasons. First, model inference and history aggregation are not naturally deterministic across peers: they may depend on floating-point libraries, sliding windows, and changing request histories. Placing them inside chaincode would violate Fabric's requirement that identical proposals yield identical read--write sets across all endorsers. Second, IoT deployments often need to update the risk model or recalibrate thresholds as the environment evolves; keeping inference off-chain allows these updates without chaincode redeployment or consortium-wide governance approval. Third, the blockchain layer should remain auditable and simple---peers only validate a signature, context fields, freshness, and replay status. These checks are compact and reproducible, making them well-suited for endorsement simulation under Fabric's execute--order--validate model.

The use of per-key SBE also provides finer granularity than a single chaincode-level endorsement policy. A chaincode-level policy applies the same endorsement requirement to all writes in a namespace, which is conservative but inefficient for mixed-risk IoT traffic. Per-key or per-key-class SBE allows benign telemetry updates to remain lightweight while sensitive or suspicious operations require broader organizational agreement. In practice, operators define templates (Relaxed, Moderate, Strict) once through consortium governance, and the oracle's tier label selects among them at runtime. This avoids redesigning consensus while making validation depth directly responsive to per-request security posture.

\subsection{Security Considerations}
\label{sec:security}

\noindent\textbf{Threat model.} We consider adversaries controlling one or more IoT clients that submit arbitrary Fabric proposals, including attempts to (i) perform unauthorized reads/writes, (ii) replay previously observed attestations, and (iii) generate high-rate bursts that stress endorsers and the orderer. An adversary may also compromise a subset of endorser nodes within a single organization, but cannot compromise a quorum of organizations required by the \emph{Strict} endorsement profile. We assume TLS-protected channels between the SDK and peers, MSP-governed identities, and correct Raft behavior under the consortium governance model. Our goals are: (a) higher-risk requests are forced through stronger cross-organization endorsement before commit; (b) replayed or stale attestations are rejected deterministically; and (c) oracle decisions remain auditable on-chain via signed labels and tier-conditioned policies. We do not aim to eliminate volumetric network-layer DoS; rather, we allocate more validation effort to suspicious operations while keeping benign operations lightweight.

\noindent\textbf{Security properties.} Given attested tier $T$, Fabric validates endorsements against the effective policy at commit time; a high-risk request cannot commit without satisfying the strict profile, even if the client targets a weaker set of peers. Freshness and replay protection follow from attestation expiry and the context-bound $\texttt{att\_id}$: stale epochs and spent identifiers are rejected deterministically, and replaying a label against a different key or identity fails the binding checks. Policy downgrade is prevented by Invariant~A, ensuring each write is validated under a single, committed endorsement rule. If the oracle signing key is compromised, requests may be mislabeled; key protection, rotation, and threshold/multiple-oracle signing are important operational extensions.

\section{Evaluation}
\label{sec:evaluation}

\subsection{Evaluation Metrics}

\noindent\textbf{Oracle-side (classifier):} We report accuracy, AUROC, AUPRC, precision, recall, and F1 score at the deployed threshold $\tau=0.001$, chosen to aggressively suppress false alarms under class imbalance while maintaining perfect recall on the test split.

\noindent\textbf{System:} We report end-to-end confirmation latency $T_{\mathrm{final}}$ (submit$\rightarrow$commit, median) and peer-side host overhead (CPU utilization and peak RSS) as the request rate increases from 2 to 12~txn/s. Latency is decomposed across endorsement, ordering, and validation/commit stages; the ordering and commit contributions remain governed by the static Raft batching configuration and vary little across tiers.

\subsection{Dataset, Testbed, and Baselines}

We train the LSTM oracle on the DS2OS traffic-traces dataset (357,952 IIoT request records; one normal class and seven attack types collapsed into a binary label)~\cite{IEEEhowto:Pahl}. A strictly time-ordered split prevents temporal leakage; the test set has prevalence $\pi \approx 2.8\%$. Features are constructed via sliding-window sequences with categorical embedding/one-hot encoding and z-score standardization. Training uses class-weighted binary cross-entropy with early stopping.

The Fabric testbed (Table~\ref{tab:fabric-platform}) uses 3 organizations with 2 peers each, CouchDB world state, TLS transport, deterministic Go/Node chaincode, and a static Raft ordering configuration. We vary only application-level tier thresholds and SBE profiles. Baselines: (i) \textbf{static endorsement}—fixed policy independent of oracle labels; (ii) \textbf{tier-conditioned SBE} (proposed); (iii) \textbf{always-strict}—all requests use the strictest profile.

\begin{table}[h!]
\caption{Blockchain testbed configuration.}
\label{tab:fabric-platform}
\begin{center}
\scalebox{0.9}{
\begin{tabular}{cl}
\hline
\textbf{Component} & \textbf{Setting} \\
\hline
Organizations & 3 orgs, 2 peers/org (endorsing/committing) \\
Orderer        & Raft (channel-static batching) \\
World State    & CouchDB (per peer) \\
Batch Timeout  & 1\,s (fixed per channel) \\
Max Msg/Block  & 50 (fixed per channel) \\
Chaincode      & Go/Node; deterministic + label verify \\
Transport      & TLS enabled; client SDK submits proposals \\
\hline
\end{tabular}}
\end{center}
\end{table}

\section{Results}
\label{sec:results}

\noindent\textbf{Confirmation latency.}
Table~\ref{tab:overall-tier} shows that higher-risk tiers incur higher confirmation latency due to stricter endorsement requirements. The ordering and commit stages remain governed by fixed Fabric parameters and vary little across tiers.

\begin{table}[h]
\centering
\caption{Median confirmation latency per tier (374 runs).}
\label{tab:overall-tier}
\setlength{\tabcolsep}{6pt}
\begin{tabular}{lcr}
\toprule
\textbf{Tier} & \textbf{SBE profile} & \textbf{Median latency (s)}\\
\midrule
Low      & Relaxed  & 0.0412 \\
Moderate & Moderate & 0.0457 \\
High     & Strict   & 0.0838 \\
\bottomrule
\end{tabular}
\end{table}

\begin{figure}[h]
  \centering
  \incfig{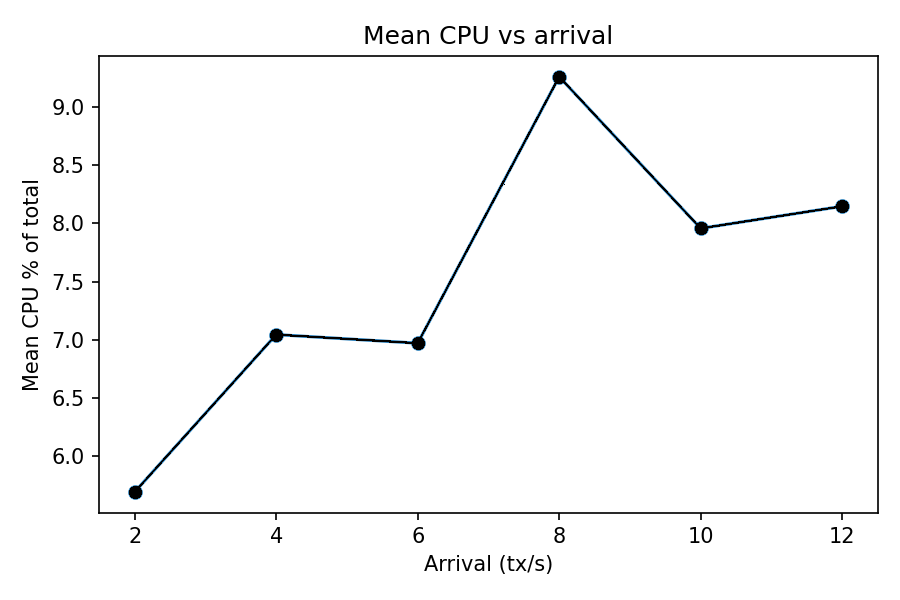}{.45\textwidth}\\[4pt]
  \incfig{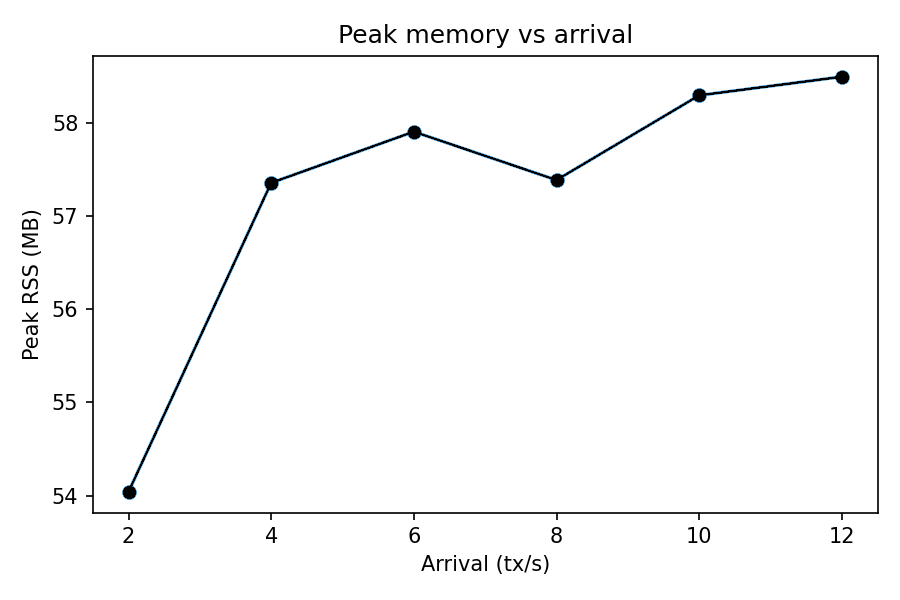}{.45\textwidth}
  \caption{Host overhead vs.\ arrival rate: mean CPU (top) and peak RSS (bottom).}
  \label{fig:resource_plots}
\end{figure}

\noindent\textbf{Host overhead.}
Figure~\ref{fig:resource_plots} shows peer-side CPU and RSS trends as arrival rate increases. On-chain verification adds modest overhead relative to baseline Fabric processing, since verifying an oracle signature and replay checks is lightweight compared to endorsement execution and commit.

Table~\ref{tab:resource_summary} reports CPU/RSS aggregates across arrival rates. Peer resource usage remains stable, confirming that on-chain verification logic introduces modest overhead.

\begin{table}[h!]
  \centering
  \caption{Resource summary across arrival rates (\texttt{bs=100}, \texttt{bt=0.2s}).}
  \label{tab:resource_summary}
  \setlength{\tabcolsep}{6pt}
  \begin{tabular}{lccc}
    \toprule
    Arrival (txn/s) & CPU mean (\%) & Peak RSS (MB) & Wall (s) \\
    \midrule
    2.0  & 45.5 & 54.04 & 1.55 \\
    4.0  & 56.4 & 57.36 & 2.06 \\
    6.0  & 55.8 & 57.91 & 2.05 \\
    8.0  & 74.1 & 57.39 & 2.07 \\
    10.0 & 63.6 & 58.30 & 2.58 \\
    12.0 & 65.2 & 58.50 & 2.61 \\
    \bottomrule
  \end{tabular}
\end{table}

\subsection{Oracle-side Detection Quality}
\label{subsec:oracle-quality}

\begin{figure}[t]
  \centering
  \begin{subfigure}[b]{0.48\columnwidth}
    \centering
    \includegraphics[width=\columnwidth]{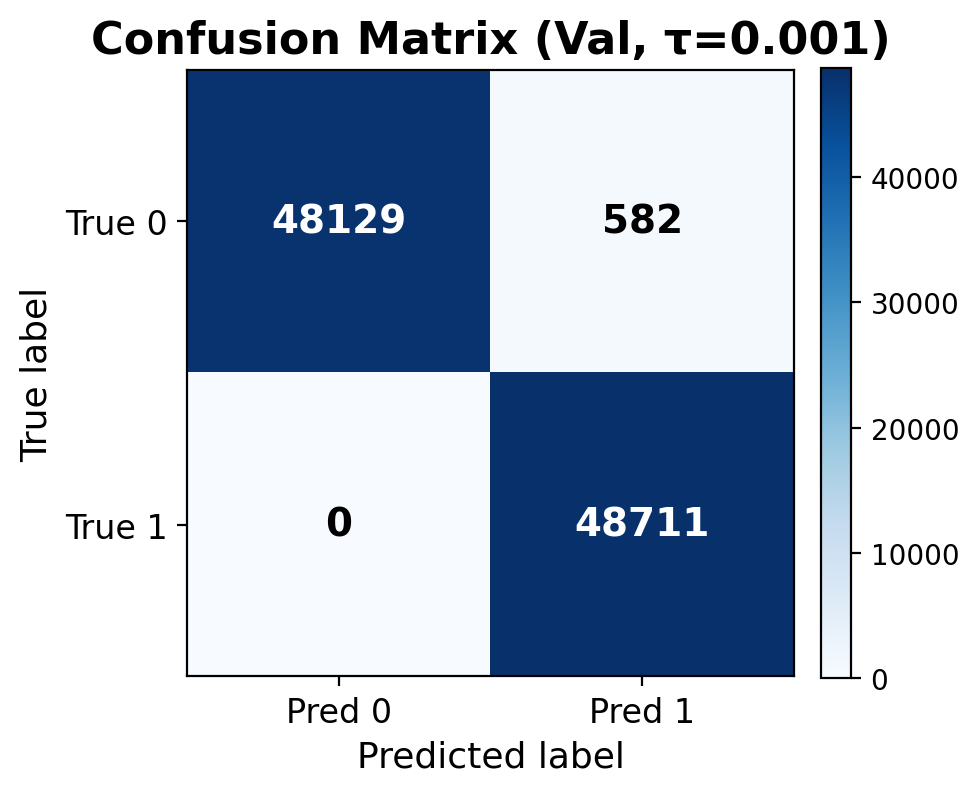}
    \caption{Validation ($\tau=0.001$).}
  \end{subfigure}\hfill
  \begin{subfigure}[b]{0.48\columnwidth}
    \centering
    \includegraphics[width=\columnwidth]{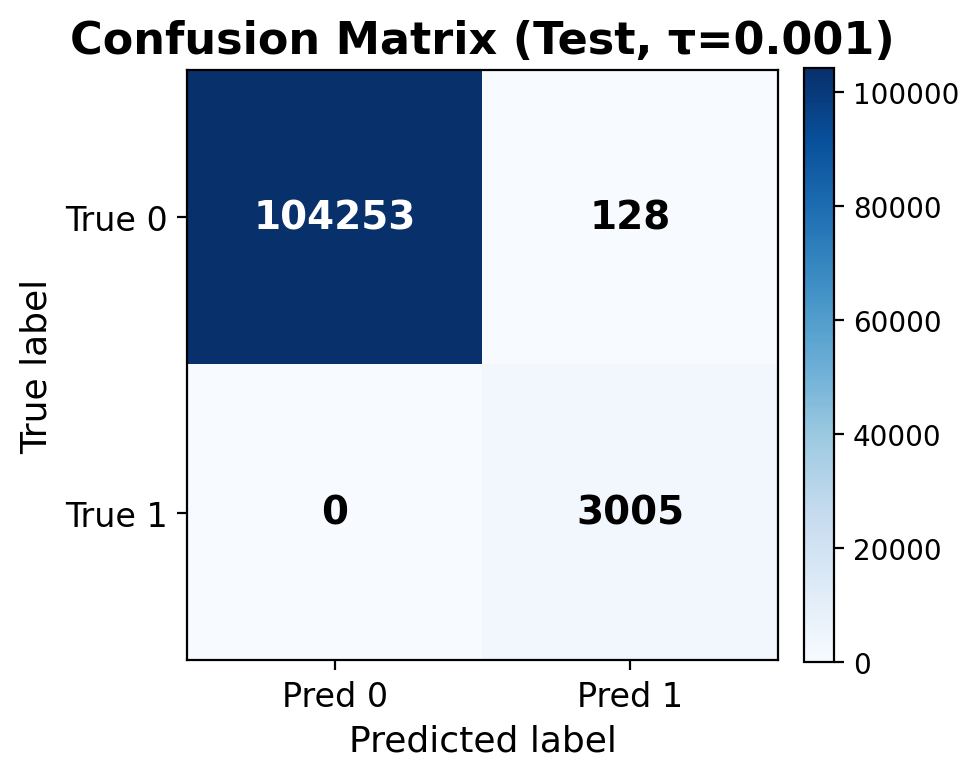}
    \caption{Test ($\tau=0.001$).}
  \end{subfigure}
  \caption{Confusion matrices at the deployed threshold.}
  \label{fig:cm-val-test}
\end{figure}

Figure~\ref{fig:cm-val-test} reports confusion matrices at the deployed threshold $\tau=0.001$. The threshold preserves perfect recall while substantially reducing false alarms on the imbalanced test set (FP reduced to 128, a ${\sim}9.5\times$ reduction without missing any attack).

\begin{table}[h!]
  \centering
  \caption{Oracle classification metrics at $\tau=0.001$.}
  \label{tab:metrics-at-tau}
  \renewcommand{\arraystretch}{1.1}
  \setlength{\tabcolsep}{8pt}
  \begin{tabular}{|l|c|c|}
    \hline
    \textbf{Metric} & \textbf{Train} & \textbf{Test} \\
    \hline
    Accuracy  & 0.9940 & 0.9988 \\
    Precision & 0.9882 & 0.9591 \\
    Recall    & 1.0000 & 1.0000 \\
    F1 score  & 0.9941 & 0.9791 \\
    AUROC     & 0.9993 & 0.9994 \\
    AUPRC     & 0.9987 & 0.9645 \\
    \hline
  \end{tabular}
\end{table}

\subsection{Sensitivity to Tier Distribution and Policy Profiles}
\label{subsec:sensitivity}

The benefit of tier-conditioned SBE is most visible under mixed-risk traffic, where benign and suspicious operations coexist in the same channel. If the \emph{Strict} profile is applied to every request, the system obtains strong validation but pays the high-risk endorsement cost for routine telemetry. If the \emph{Relaxed} profile is applied universally, latency remains low but suspicious writes receive no additional organizational scrutiny. The proposed mapping separates these cases: routine traffic stays on the relaxed path, while suspicious traffic is shifted to profiles requiring additional peers or broader organization diversity.

This separation also provides a deployment knob for administrators. The thresholds that produce \Low, \Mod, and \High labels can be tuned independently from the endorsement templates. For example, a safety-critical actuator namespace may map \Mod requests directly to a stricter cross-organization policy, while a low-impact sensing namespace may reserve the strict profile only for \High requests. The same oracle output can thus support different security postures across key classes without modifying the orderer.

The results should therefore be interpreted as a \emph{validation-cost allocation} strategy rather than a replacement for admission control. Network-level rate limiting and gateway filtering reduce volumetric pressure before requests reach Fabric, whereas tier-conditioned SBE determines how much independent validation is required once a transaction is processed by the ledger. Combining these layers is important for IoT deployments: admission controls protect availability, while risk-adaptive endorsement improves integrity and accountability for state-changing operations. The committed transaction history also supports post-incident analysis: operators can inspect the attested tier, the endorsement profile used, and whether threshold changes affected the distribution of \Low, \Mod, and \High decisions over time.

\subsection{Comparison with Related Mechanisms}
\label{subsec:capability-matrix}

Table~\ref{tab:cap-matrix} compares representative mechanisms across four properties: \emph{risk-adaptive admission} (controls before transaction submission), \emph{risk-adaptive validation} (stricter authorization during execution), \emph{Fabric-native} (no orderer/consensus changes), and \emph{deterministic on-chain} verification. The proposed approach is the only design satisfying all four.

\begin{table}[h!]
\centering
\caption{Capability matrix.}
\label{tab:cap-matrix}
\setlength{\tabcolsep}{3pt}
\renewcommand{\arraystretch}{1.15}
\scriptsize
\resizebox{\columnwidth}{!}{%
\begin{tabular}{lcccc}
\toprule
\textbf{Approach} &
\textbf{\begin{tabular}[c]{@{}c@{}}Risk-adaptive\\Admission\end{tabular}} &
\textbf{\begin{tabular}[c]{@{}c@{}}Risk-adaptive\\Validation\end{tabular}} &
\textbf{\begin{tabular}[c]{@{}c@{}}Fabric-native\\(no orderer changes)\end{tabular}} &
\textbf{\begin{tabular}[c]{@{}c@{}}Deterministic\\On-chain\end{tabular}} \\
\midrule
\textbf{Proposed} & \checkmark & \checkmark & \checkmark & \checkmark\\
Bobde~\cite{bobde}& $\times$ & $\times$ & $\times$ & $\times$\\
Saha~\cite{saha}  & $\times$ & $\times$ & $\times$ & \checkmark\\
Hu~\cite{Hu}      & $\times$ & $\times$ & $\times$ & $\times$\\
\bottomrule
\end{tabular}%
}
\end{table}

\section{Discussion and Limitations}
\label{sec:discussion}

Tier separation is primarily governed by endorsement strictness, while the Fabric ordering/commit path remains stable across tiers under a fixed network configuration. In deployments, tier thresholds should be calibrated to the expected base rate of suspicious events and the cost of false positives: start conservative (keeping most benign traffic in \Low), monitor per-tier endorsement load and latency, then adjust jointly. SBE profile templates (Relaxed/Moderate/Strict) can be predefined through consortium governance and kept stable, with the oracle activating the appropriate template per request. A sudden increase in \High-tier labels may indicate either a real attack burst or a model calibration problem, both of which should trigger operational review. Similarly, operators can compare the distribution of tier labels and confirmation latency before and after threshold recalibration to measure the impact of policy changes quantitatively.

\noindent\textbf{Operational notes.}
Oracle key management requires distribution via channel configuration and rotation with overlap periods. Spent-registry growth should be bounded via epoch-based eviction or per-epoch bloom filters. In high-churn workloads, coarse-grained (per-namespace) policies reduce SBE update frequency. Because SBE profile semantics are known in advance and only the oracle's tier label determines which template is activated, the system remains easy to audit: policy logic is stable while per-request risk drives template selection.

\noindent\textbf{Limitations.}
(1) The oracle is trusted to sign correct tiers; compromised keys can mislabel requests. (2) Tiering quality depends on training data and threshold calibration; distribution shifts may increase false positives/negatives. (3) We do not model adversaries that collude across endorsing organizations or manipulate off-chain metadata streams. (4) Frequent tier changes increase metadata-only SBE update transactions and should be bounded by rate limits. (5) Request metadata may be sensitive; privacy-preserving feature extraction remains an important direction for future work.

\section{Conclusion}

This paper introduced a risk-adaptive IoT access-control framework combining off-chain learning with deterministic on-chain enforcement on Hyperledger Fabric. An LSTM oracle maps each request to \textsf{Low}/\textsf{Moderate}/\textsf{High} tiers and issues a signed attestation bound to the submitting client and target key/version. Endorsing peers verify the attestation deterministically in chaincode and enforce a tier-conditioned SBE profile without modifying the ordering service. Experiments show stricter endorsement profiles harden validation for higher-risk requests while keeping low-risk confirmation latency near the baseline. The audit trail embedded in each committed transaction further supports post-incident analysis and threshold recalibration over time. Future work includes threshold oracle signing, privacy-preserving feature extraction, and evaluation under adversarial traffic distributions, offering a practical path toward risk-adaptive security controls in permissioned blockchain-based IoT systems.

\end{document}